\documentclass{article}

\usepackage[preprint]{neurips_2024}   
\makeatletter
\renewcommand{\@noticestring}{Preprint. arXiv submission.}
\makeatother
\usepackage[utf8]{inputenc}
\usepackage[T1]{fontenc}
\usepackage{hyperref}
\usepackage{url}
\usepackage{booktabs}
\usepackage{amsfonts}
\usepackage{amsmath}
\usepackage{amssymb}
\usepackage{nicefrac}
\usepackage{microtype}
\usepackage{xcolor}
\usepackage{graphicx}
\usepackage{multirow}
\usepackage{listings}
\title{Dynamic Attentional Context Scoping:\\
Agent-Triggered Focus Sessions for\\
Isolated Per-Agent Steering in Multi-Agent LLM Orchestration}

\author{%
  Nickson Patel \\
  Independent Researcher \\
  \texttt{nicksonpatel@gmail.com}
}

\begin{document}
\maketitle
\renewcommand{\arraystretch}{1.15}

\begin{abstract}
Multi-agent LLM orchestration systems suffer from \emph{context pollution}:
when $N$ concurrent agents compete for the orchestrator's context window,
each agent's task state, partial outputs, and pending questions contaminate
the steering interactions of every other agent, degrading decision quality.
We introduce \textbf{Dynamic Attentional Context Scoping (DACS)}, a
mechanism in which the orchestrator operates in two asymmetric modes.
In \textsc{Registry} mode it holds only lightweight per-agent status
summaries ($\leq 200$ tokens each), remaining responsive to all agents
and the user.
When an agent emits a \textsc{SteeringRequest}, the orchestrator enters
\textsc{Focus}$(a_i)$ mode, injecting the full context of agent $a_i$
while compressing all other agents to their registry entries.
Context isolation is agent-triggered, asymmetric, and deterministic:
the context window contains exactly $F(a_i) + R_{-i}$ during steering,
eliminating cross-agent contamination without requiring context
compression or retrieval.
We evaluate DACS across four experimental phases totalling 200 trials:
Phase~1 tests $N \in \{3,5,10\}$ agents (60 trials);
Phase~2 tests agent heterogeneity and adversarial dependency structures
(60 trials);
Phase~3 tests decision density scaling up to $D=15$ steering requests per
agent (40 trials);
Phase~4 replaces all scripted stubs with autonomous LLM agents generating
free-form questions (40 trials, Claude Haiku~4.5).
Across all 8 synthetic scenarios, DACS achieves $90.0$--$98.4\%$ steering
accuracy versus $21.0$--$60.0\%$ for a flat-context baseline
($p < 0.0001$ throughout), with wrong-agent contamination falling from
$28$--$57\%$ to $0$--$14\%$ and context efficiency ratios of up to
$3.53\times$.
The accuracy advantage grows with both agent count ($N$) and decision
density ($D$), and keyword matching is validated by LLM-as-judge across
all three synthetic phases (mean $\kappa=0.909$).
In Phase~4, all scripted stubs are replaced by autonomous LLM agents
(Claude Haiku~4.5) generating free-form steering questions:
DACS outperforms the flat-context baseline by $+17.2$\,pp at $N{=}3$
($p{=}0.0023$) and $+20.4$\,pp at $N{=}5$ ($p{=}0.0008$),
with the advantage growing with $N$ confirmed by two independent judges.
All code and data are released at
\href{https://github.com/nicksonpatel/dacs-agent-focus-mode}{\texttt{github.com/nicksonpatel/dacs-agent-focus-mode}}.
\end{abstract}

\section{Introduction}
\label{sec:intro}

The deployment of LLMs as orchestrators managing multiple parallel agents
has become a practical reality: systems such as Claude Code Agent
Teams~\citep{anthropic2025code}, OpenCode~\citep{opencode2025}, and
production multi-agent platforms~\citep{agentorchestra2025} demonstrate
that complex long-horizon tasks can be decomposed across concurrent
specialized agents.
The common architectural choice is a single LLM instance (the
\emph{orchestrator}) that coordinates all agents, handles user
interaction, and steers individual agents when they reach decision points.

\paragraph{The context pollution problem.}
This architecture introduces a fundamental scaling problem.
When $N$ agents run concurrently, each maintains its own task context:
partial outputs, domain-specific state, pending decisions.
In a flat-context orchestrator, all of these threads compete for space in
a single context window.
When agent $a_i$ requests steering—``should I use BFS or DFS for this
graph traversal?''—the orchestrator's context simultaneously holds $a_j$'s
transformer attention survey, $a_k$'s CSV encoding problem, and so on.
We call this \emph{context pollution}: the systematic contamination of
one agent's steering interaction by irrelevant context from other agents.

The consequences are measurable and severe.
Prior work has documented context pollution in single-agent settings
under tool overload~\citep{adaptorch2025,codedelegator2025}, under
retrieval bloat~\citep{sidequest2025}, and under long conversation
histories~\citep{afm2024,ace2026}.
Our experiments show that in the multi-agent setting, a flat-context
orchestrator's steering accuracy collapses from 60\% at $N=3$ to 21\% at
$N=10$ (Phase~1), and further degrades as agents become more diverse
(Phase~2) or accumulate longer decision histories (Phase~3).

\paragraph{The DACS mechanism.}
We propose \textbf{Dynamic Attentional Context Scoping (DACS)}, which
solves context pollution through \emph{agent-triggered asymmetric context
isolation}.
DACS introduces two orchestrator modes:

\begin{itemize}
  \item \textbf{\textsc{Registry} mode.} The orchestrator holds only the
  registry $R = \{r_1, \ldots, r_N\}$, one compact status summary per
  agent ($\leq 200$ tokens each).
  It monitors all agents, responds to the user, and queues incoming
  steering requests.

  \item \textbf{\textsc{Focus}$(a_i)$ mode.} When agent $a_i$ emits a
  \textsc{SteeringRequest}, the orchestrator transitions to hold $F(a_i)$
  (the full context of $a_i$: task description, prior steering exchanges,
  current partial output) plus a compressed version of $R_{-i}$
  (registry summaries for all agents except $a_i$).
  The context window contains exactly what is needed to steer $a_i$,
  and nothing from any other agent's task thread.
\end{itemize}

The key properties of DACS that distinguish it from prior context
management approaches are:
\begin{enumerate}
  \item \textbf{Agent-triggered.} Context isolation fires when an agent
  needs it, not on every turn or at task-decomposition time.
  \item \textbf{Asymmetric.} The steered agent gets full fidelity;
  all others get compressed summaries.
  \item \textbf{Exact, not approximate.} No scoring function, no fidelity
  tiering~\citep{afm2024,lemon2025}, no probabilistic compression: the
  context is deterministically constructed as $F(a_i) + R_{-i}$.
  \item \textbf{Sub-linear in $N$ and $D$.} Focus context size is
  independent of the number of other agents; only the compressed registry
  grows with $N$ (by $\approx 25$ tokens per additional agent in our
  experiments). Accuracy is also stable as decision density $D$ increases.
\end{enumerate}

\paragraph{Contributions.}
\begin{enumerate}
  \item We formally define the DACS mechanism: the \textsc{Registry}/
  \textsc{Focus} state machine, the \textsc{SteeringRequest} protocol,
  and the context builder invariants (\S\ref{sec:mechanism}).
  \item We implement a minimal, fully observable orchestration harness
  ($\sim$300 lines) where every token entering each LLM call is logged,
  enabling strict experimental control (\S\ref{sec:experiments}).
  \item We show large, statistically significant accuracy gains across
  Phase~1 ($N$ scaling), Phase~2 (agent diversity and adversarial
  dependencies), and Phase~3 (decision density scaling), together
  answering four distinct research questions (\S\ref{sec:results}).
  \item We independently validate the keyword matching metric with
  LLM-as-judge evaluations across Phases~2 and~3 (Phase~2: $\kappa=0.956$;
  Phase~3: mean $\kappa=0.909$), confirming that reported advantage sizes
  are robust (\S\ref{sec:results}).
  \item We validate beyond the scripted benchmark with Phase~4: two
  real-agent experiments (40 trials, Claude Haiku~4.5, free-form questions)
  showing DACS outperforms the baseline by $+17.2$\,pp and $+20.4$\,pp at
  $N{=}3$ and $N{=}5$ respectively, with the advantage growing with $N$
  (\S\ref{sec:phase4}).
  \item We identify sub-linear context scaling as the mechanism's key
  theoretical property, explain its source, and discuss conditions under
  which it holds (\S\ref{sec:discussion}).
\end{enumerate}

\section{Related Work}
\label{sec:related}

Context management for LLM agents has been studied at three levels of
granularity: single-agent memory across turns, single-agent tool/retrieval
bloat, and multi-agent orchestration.
DACS addresses a problem at the third level that existing work at the
first two levels does not reach.

\paragraph{Single-turn context compression.}
\begin{sloppypar}
AFM~\citep{afm2024} assigns per-message fidelity tiers
(Full/Compressed/Placeholder) based on a composite recency-relevance score
and packs messages greedily under a token budget.
It achieves 83.3\% constraint-recall on its benchmark, validating that
structured fidelity management outperforms naive truncation.
ACE~\citep{ace2026} treats the agent's context as an evolving playbook
of itemized bullets with utility counters, using incremental delta updates
to avoid context collapse (they observe a 18{,}282-token context collapsing
to 122 tokens under iterative rewriting, dropping accuracy below the
no-adaptation baseline).
Both AFM and ACE operate within a single agent's context history.
Neither addresses $N$ concurrent agents competing for an orchestrator
context window, and neither models wrong-agent contamination as a failure
mode.
\end{sloppypar}

\paragraph{Retrieval and cache bloat.}
SideQuest~\citep{sidequest2025} frames KV cache compression as an
auxiliary task solved by the LRM itself, achieving 65\% peak token
reduction on single-agent deep research tasks.
Lemon Agent~\citep{lemon2025} applies three-tier progressive compression
(full → compressed → summary) to a shared orchestrator context over task
time in an orchestrator-worker system.
Both treat context as a resource to compress; neither introduces asymmetric
isolation triggered by an agent's request.

\paragraph{Context pollution in tool-heavy agents.}
Adaptive Orchestration~\citep{adaptorch2025} identifies context pollution
and attention decay as failure modes in monolithic agents loaded with too
many tools, and proposes spawning specialist sub-agents (DMoE) to offload
capability.
CodeDelegator~\citep{codedelegator2025} separates planner and executor
roles via Ephemeral-Persistent State Separation (EPSS): a persistent
Delegator never receives execution traces; each Coder sub-task starts with
a clean context.
Both solve context pollution \emph{architecturally} (at task decomposition
time) for sequential delegation to one sub-agent at a time.
DACS solves it \emph{dynamically at runtime}, during the steering
interaction itself, for $N$ agents running simultaneously.
The distinction matters: CodeDelegator's Delegator has no mechanism for
handling the case where multiple concurrent Coders simultaneously need
steering.

\paragraph{Multi-agent orchestration context management.}
AOI~\citep{aoi2024} introduces a three-layer memory architecture for
multi-agent IT operations, with a central Context Compressor transforming
raw agent outputs into a compressed cache.
Its Observer agent always holds a compressed aggregate of \emph{all}
agent activity simultaneously; there is no per-agent isolation.
Agents execute and report results; they cannot signal the orchestrator
for steering attention, and wrong-agent contamination is not modelled
(AOI's failure mode is information loss from operational data volume).
AgentOrchestra~\citep{agentorchestra2025} achieves SOTA on GAIA (89.04\%)
via hierarchical delegation: a planning agent routes sub-tasks to
domain-specific sub-agents, bounding the planner's context footprint by
converting global coordination into localized routing decisions.
The paper explicitly notes that flat coordination ``tends to accumulate
irrelevant context''—independent validation of the problem DACS solves.
However, hierarchical routing is a structural (pre-execution) partial
solution: it reduces \emph{how much} context enters the orchestrator, but
agents still request the orchestrator's attention concurrently at runtime,
and the orchestrator has one flat context window for those interactions.
DACS and AgentOrchestra are complementary: AgentOrchestra reduces the
initial context volume; DACS controls what the orchestrator holds at the
moment of each steering decision.
AdaptOrch~\citep{adaptorch2025b} formalizes topology selection as the
dominant performance variable as frontier LLMs converge, showing
$+6.9$--$9.8$pp gains on SWE-bench/GPQA/HotpotQA from topology-aware
routing.
AdaptOrch makes one topology decision per task; DACS makes repeated
context-switch decisions throughout execution.

\paragraph{Summary.}
No prior work implements agent-triggered asymmetric \textsc{Registry}/
\textsc{Focus} mode switching for per-agent context isolation in
concurrent multi-agent orchestration.
Table~\ref{tab:related} summarises the key distinctions.

\begin{table}[t]
\caption{Comparison of context management mechanisms.}
\label{tab:related}
\centering
\small
\begin{tabular}{lcccc}
\toprule
\textbf{System} & \textbf{$N$ concurrent} & \textbf{Asym.\ mode switch} & \textbf{Agent-triggered} & \textbf{Contamination metric} \\
\midrule
AFM              & \texttimes & \texttimes & \texttimes & \texttimes \\
ACE              & \texttimes & \texttimes & \texttimes & \texttimes \\
AOI              & \checkmark & \texttimes & \texttimes & \texttimes \\
AgentOrchestra   & \checkmark & \texttimes & \texttimes & \texttimes \\
AdaptOrch        & \checkmark & \texttimes & \texttimes & \texttimes \\
CodeDelegator    & \texttimes & \texttimes & \texttimes & \texttimes \\
Adaptive Orch.   & \texttimes & \texttimes & \texttimes & \texttimes \\
Lemon Agent      & \checkmark & \texttimes & \texttimes & \texttimes \\
\midrule
\textbf{DACS (ours)} & \checkmark & \checkmark & \checkmark & \checkmark \\
\bottomrule
\end{tabular}
\end{table}

\section{Dynamic Attentional Context Scoping (DACS)}
\label{sec:mechanism}

\subsection{Entities and Notation}
\label{sec:notation}

Let $O$ denote the orchestrator LLM, $A = \{a_1, \ldots, a_N\}$ the set
of $N$ concurrently running agents, and $T$ the context window token
budget.
Each agent $a_i$ has an associated \emph{focus context}
$F(a_i)$, the full set of information the orchestrator needs to steer
$a_i$: task description, previous steering exchanges between $O$ and
$a_i$, and $a_i$'s current partial output summary.

The \emph{registry} $R = \{r_1, \ldots, r_N\}$ is a set of compact status
snapshots, one per agent.
Each entry $r_i$ contains:

\begin{center}
\begin{tabular}{ll}
\texttt{agent\_id} & identifier \\
\texttt{task} & task description ($\leq 50$ tokens) \\
\texttt{status} & \textsc{Running} $|$ \textsc{Blocked} $|$ \textsc{Waiting} $|$ \textsc{Complete} \\
\texttt{last\_output\_summary} & $\leq 100$ tokens \\
\texttt{urgency} & \textsc{Low} $|$ \textsc{Medium} $|$ \textsc{High} \\
\end{tabular}
\end{center}

Target budget per entry: $\leq 200$ tokens.
Full registry size for $N=10$: $\leq 2{,}000$ tokens, leaving ample space for $F(a_i)$.

\subsection{Orchestrator State Machine}
\label{sec:fsm}

The orchestrator operates as an explicit finite-state machine with three states:

\begin{description}
  \item[\textsc{Registry}.] $O$ holds $R$ only.
  It monitors all agents, responds to user messages, and processes
  incoming \textsc{SteeringRequest}s from the queue.

  \item[\textsc{Focus}$(a_i)$.] $O$ holds $F(a_i)$ and a compressed view
  of $R_{-i}$ (all registry entries except $a_i$'s, since $a_i$'s full
  context is already in $F(a_i)$).
  $O$ steers $a_i$ and cannot accept new steering requests until the
  focus session ends.

  \item[\textsc{UserInteract}.] $O$ responds to a user message using $R$
  only (same as \textsc{Registry} mode, but explicitly user-facing).
  Queued steering requests are not processed until this state exits.
\end{description}

\noindent\textbf{Transitions:}
\begin{align}
  \textsc{Registry} &\;\to\; \textsc{Focus}(a_i): \quad
    a_i \text{ emits } \textsc{SteeringRequest}(\text{urgency} = \text{any}) \label{eq:t1}\\
  \textsc{Focus}(a_i) &\;\to\; \textsc{Registry}: \quad
    \textsc{SteeringComplete} \text{ or } \textsc{SteeringAbandoned} \label{eq:t2}\\
  \textsc{Focus}(a_i) &\;\to\; \textsc{Focus}(a_j): \quad
    a_j \text{ emits } \textsc{SteeringRequest}(\text{urgency} = \textsc{High}), \; a_j \neq a_i \label{eq:t3}\\
  \textsc{Registry} &\;\to\; \textsc{UserInteract}: \quad
    \text{user message received} \label{eq:t4}
\end{align}

Transition~\eqref{eq:t3} (the \emph{interrupt} protocol) allows a
high-urgency agent to preempt an active focus session.
On interrupt, the current partial steering state for $a_i$ is saved and
a new focus context for $a_j$ is built.
After $a_j$ is steered, the orchestrator returns to $\textsc{Registry}$
and resumes $a_i$'s interrupted session if it is still pending.

\subsection{The SteeringRequest Protocol}
\label{sec:protocol}

When agent $a_i$ reaches a decision point it cannot resolve autonomously,
it emits:

\begin{lstlisting}
SteeringRequest {
  agent_id:   str
  context:    str   -- relevant context excerpt
  question:   str   -- specific decision needed
  blocking:   bool  -- is a_i halted?
  urgency:    LOW | MEDIUM | HIGH
}
\end{lstlisting}

\textsc{High} urgency can interrupt an active focus session
(transition~\eqref{eq:t3}).
\textsc{Medium} requests queue; the agent continues on a default path.
\textsc{Low} requests are batched.

\subsection{Context Builder}
\label{sec:context_builder}

The context builder constructs the orchestrator's prompt before each LLM
call.
It enforces a hard token budget $T$ deterministically (token count is
checked and enforced before every call; the LLM provider is never relied
upon for truncation).

\begin{description}
  \item[\texttt{build\_focus\_context}$(a_i)$.]
  Returns $F(a_i) \,\|\, \texttt{compress}(R_{-i})$.
  If $|F(a_i)| + |R_{-i}| > T$, registry entries for lower-urgency
  agents are progressively truncated until the budget is met.
  $a_i$'s focus context $F(a_i)$ is never truncated.

  \item[\texttt{build\_registry\_context}$()$.]
  Returns $R$ in full.
  Used during \textsc{Registry} and \textsc{UserInteract} modes.
\end{description}

\noindent\textbf{Invariants enforced by the context builder:}
\begin{enumerate}
  \item $O$ is never simultaneously in \textsc{Focus} mode for more than
  one agent. \label{inv:1}
  \item User messages are never dropped; they are queued during
  \textsc{Focus} and processed on the next \textsc{Registry} entry.
  \item $R$ is always current within the last agent heartbeat (agents push
  a compact status update after each step). \label{inv:3}
  \item $|C| \leq T$ holds before every LLM call without exception. \label{inv:4}
\end{enumerate}

\subsection{Complexity Analysis}
\label{sec:complexity}

Let $|F|$ denote the average focus context size (task-dependent, independent
of $N$) and $|r|$ the average registry entry size ($\leq 200$ tokens).

\paragraph{\textsc{Registry} mode context size.}
$|C_{\text{reg}}| = N \cdot |r|$, growing linearly in $N$.
However, this mode is lightweight: the orchestrator makes no costly
steering decisions here, only monitoring and routing.

\paragraph{\textsc{Focus}$(a_i)$ mode context size.}
$|C_{\text{focus}}| = |F| + (N-1) \cdot |r|$.
The focus context $|F|$ is independent of $N$.
Only the compressed registry of $N-1$ agents adds $O(N)$ tokens.
For large $|F|$ and small $|r|$ (our target design point), this is
approximately $|F| + O(N \cdot |r|)$ where $|r| \ll |F|$.

In Phase~1 experiments, $|F| \approx 450$--$550$ tokens per agent and
$|r| \approx 25$ tokens, giving $|C_{\text{focus}}| \approx 500 + 25N$,
empirically confirmed: 561 tokens at $N=3$, 633 at $N=5$, 816 at $N=10$
($+25.5$ tokens/agent, $R^2 > 0.99$).

In Phase~3 (high decision density, $D=15$), $|F|$ grows with the
accumulated steering history, yielding $|C_{\text{focus}}| \approx 2{,}755$
tokens, higher than Phase~1 in absolute terms, but isolated to exactly
$a_i$'s history.
The flat baseline accumulates \emph{all} agents' histories simultaneously,
reaching 6{,}573 tokens (2.39$\times$ ratio).

\paragraph{Efficiency ratio.}
$\rho(N) = |C_{\text{flat}}| / |C_{\text{focus}}| \approx N|F| / (|F| +
N|r|)$.
As $N \to \infty$, $\rho \to |F| / |r| \approx 20$--$22\times$ at our
design point.
Empirically, $\rho$ grows from 2.12$\times$ ($N=3$) to 3.53$\times$
($N=10$) in Phase~1, and reaches 3.24$\times$ at $N=5, D=8$ in Phase~3.

\section{Experimental Setup}
\label{sec:experiments}

\subsection{Harness Design}
\label{sec:harness}

We implement a minimal Python orchestration harness ($\sim$300 lines
across \texttt{src/}) with full context observability: \emph{every} token
entering each LLM call is logged to a \texttt{.jsonl} trial file.
This is the central experimental variable.
We use no external orchestration framework (LangGraph, CrewAI, etc.);
their context assembly internals would add uncontrolled noise.

The harness has four components:
\texttt{registry.py} (per-agent state management),
\texttt{protocols.py} (\textsc{SteeringRequest/Response} dataclasses),
\texttt{context\_builder.py} (token-counted context construction),
\texttt{orchestrator.py} (state machine + LLM call dispatch).

\paragraph{LLM backend.}
Phases~1--3 use MiniMax-M2.7 via an Anthropic-compatible API endpoint
(context window 204{,}800 tokens).
Phase~4 uses Claude Haiku~4.5 via OpenRouter (agents and orchestrator both
use the same model, maintaining a single-model stack with no confound from
model differences within each phase).
Within each phase, DACS and the baseline use the identical model and endpoint.

\subsection{Conditions}
\label{sec:conditions}

\paragraph{DACS.} Orchestrator operates with full \textsc{Registry}/
\textsc{Focus} mode switching as described in \S\ref{sec:mechanism}.

\paragraph{Baseline (flat context).} All agents' full contexts are
injected simultaneously into every steering call.
The baseline uses the \emph{identical} code path as DACS, with the single
difference that \texttt{build\_focus\_context} is replaced by
\texttt{build\_flat\_context} (concatenates all $F(a_i)$).
No other differences.

\subsection{Task Suite}
\label{sec:tasks}

We design 8 scenarios across three experimental phases,
each with known-correct answers defined per decision point.
Correctness is evaluated by keyword matching: a steering response scores
correct if it contains $\geq 1$ expected ground-truth keyword for agent
$a_i$'s current decision point.
Each agent's correct-answer keywords are domain-specific and orthogonal
to all other agents; cross-agent vocabulary leakage is unambiguous.

\paragraph{Phase~1 — Agent count scaling (RQ1, RQ2).}
Three canonical scenarios vary $N \in \{3, 5, 10\}$ with $D \approx 3$
decisions per agent.

\begin{itemize}
  \item \textbf{s1\_n3} ($N=3$): Code writer (BST), research agent
  (transformer attention survey), data processor (CSV encoding).
  15 steering interactions per trial.

  \item \textbf{s2\_n5} ($N=5$): Above three plus graph algorithm agent
  (BFS/DFS) and RL survey agent.
  15 steering interactions per trial.

  \item \textbf{s3\_n10} ($N=10$): Above five plus federated learning,
  e-commerce churn, LRU cache, LLM alignment, and clinical trial
  pre-processing agents.
  30 steering interactions per trial.
\end{itemize}

\paragraph{Phase~2 — Agent diversity (RQ3).}
Three scenarios probe whether the DACS advantage generalises across
agent heterogeneity structures, including an adversarial case.

\begin{itemize}
  \item \textbf{s4\_n3\_homogeneous} ($N=3, D=4$): Three agents from the
  same domain (algorithm coding: red-black tree, open-addressing hash
  table, directed weighted graph).
  Minimal-contamination case: DACS must help even when domains share
  vocabulary.

  \item \textbf{s5\_n5\_crossfire} ($N=5, D=4$): Five maximally diverse
  domains: lock-free C++ queue, diffusion model survey, genomics VCF
  ETL, C++ memory-leak debugger, clinical trial methodology.
  Guaranteed vocabulary disjointness maximises the contamination signal.

  \item \textbf{s6\_n5\_cascade} ($N=5, D=3$): Five pipeline-dependent
  agents (planner $\to$ retrieval $\to$ ranking $\to$ feature store $\to$
  reviewer in a recommendation system).
  Adversarial for DACS: the flat baseline may benefit from seeing all
  agents' histories simultaneously when outputs depend on each other.
\end{itemize}

\paragraph{Phase~3 — Decision density scaling (RQ4).}
Two scenarios fix low $N$ but raise $D$ substantially to probe whether
DACS advantage compounds with decision history depth.

\begin{itemize}
  \item \textbf{s7\_n5\_dense\_d2} ($N=5, D=8$, 40 total steering
  interactions): Five diverse agents: async web scraper, federated
  learning survey, fraud detection pipeline, flaky-test debugger,
  distributed cache TDD.
  Doubles Phase 1's $D$ at fixed $N=5$.

  \item \textbf{s8\_n3\_dense\_d3} ($N=3, D=15$, 45 total steering
  interactions): Three agents: BERT legal-text classifier training loop,
  clinical trial hypothesis testing, post-quantum cryptography
  whitepaper.
  Quintuples Phase 1's $D$ at fixed $N=3$, the most decision-history-
  intensive scenario tested.
\end{itemize}

\subsection{Metrics}
\label{sec:metrics}

\begin{description}
  \item[\textbf{Steering accuracy.}]
  For each steering interaction, we check whether the orchestrator's
  response contains the expected ground-truth keywords for agent $a_i$.
  Score: fraction of interactions with correct keyword hit, averaged
  across all steering interactions in the trial.

  \item[\textbf{Wrong-agent contamination.}]
  For each steering response targeting $a_i$, we check whether the
  response contains ground-truth keywords for any $a_j$, $j \neq i$.
  Score: fraction of interactions with cross-agent keyword leakage.

  \item[\textbf{Context size at steering.}]
  Token count of the orchestrator's context window at the moment the
  steering LLM call is made.
  Logged directly from the context builder.
\end{description}

\paragraph{Metric validation.}
Keyword matching is validated at two stages via independent LLM-as-judge
evaluations using the same MiniMax-M2.7 model.
In Phase~2 all 400 decisions in \textbf{s5\_n5\_crossfire} were judged:
agreement 98.0\% ($\pm 1.4\%$); Cohen's $\kappa = 0.956$ (near-perfect).
For Phase~3 we ran a stratified judge pass on both dense scenarios:
100 sampled decisions from \textbf{s8\_n3\_dense\_d3} (50 DACS~$+$~50 baseline)
and 200 from \textbf{s7\_n5\_dense\_d2} (100 DACS~$+$~100 baseline).
Agreement: 95.0\% ($\kappa=0.886$) for s8; 97.0\% ($\kappa=0.933$) for s7;
mean $\kappa=0.909$ across both Phase~3 scenarios.
All four judge evaluations ($\kappa \geq 0.886$) establish keyword matching
as a valid proxy for LLM-judged correctness across the full experiment series.

\subsection{Procedure}
\label{sec:procedure}

We run 10 independent trials per condition per scenario.
Phases~1--3 span 160 trials (60 Phase~1, 60 Phase~2, 40 Phase~3);
Phase~4 adds 40 real-agent trials (200 total).
Each trial uses a fresh agent instantiation with randomised task parameter
order. Phase~3 and Phase~4 trials were run in parallel background processes
to exploit concurrent API capacity; file writes to the shared summary CSV
are protected by \texttt{fcntl} advisory locking to prevent corruption.
Phases~1--3 results are written to \texttt{results/summary.csv};
Phase~4 results to \texttt{results\_real\_agent\_haiku/};
both include per-trial \texttt{.jsonl} logs for full context-window audit.

\section{Results: Phases 1--3}
\label{sec:results}

\subsection{Phase 1: Agent Count Scaling}

\begin{table}[t]
\caption{Phase~1 results: steering accuracy, contamination, and context
size across $N \in \{3,5,10\}$.
Values are mean $\pm$ SE over 10 trials.
$\Delta$: DACS minus baseline.
Ctx ratio: baseline mean / DACS mean.}
\label{tab:phase1}
\centering
\small
\begin{tabular}{llrrrrr}
\toprule
$N$ & Cond. & Accuracy & Contamination & Ctx (tok) & $\Delta$ acc & Ctx ratio \\
\midrule
\multirow{2}{*}{3}
  & DACS     & $96.7\pm2.1\%$ & $3.3\pm3.7\%$  & $561\pm1$   & \multirow{2}{*}{$+36.7$pp} & \multirow{2}{*}{$2.12\times$} \\
  & Baseline & $60.0\pm11.5\%$ & $56.7\pm13.3\%$ & $1{,}191\pm0$ & & \\[4pt]
\multirow{2}{*}{5}
  & DACS     & $96.7\pm5.7\%$ & $14.0\pm10.6\%$ & $633\pm25$  & \multirow{2}{*}{$+58.0$pp} & \multirow{2}{*}{$2.72\times$} \\
  & Baseline & $38.7\pm10.3\%$ & $52.0\pm14.3\%$ & $1{,}720\pm151$ & & \\[4pt]
\multirow{2}{*}{10}
  & DACS     & $90.0\pm3.0\%$ & $3.7\pm5.0\%$  & $816\pm11$  & \multirow{2}{*}{$+69.0$pp} & \multirow{2}{*}{$3.53\times$} \\
  & Baseline & $21.0\pm5.0\%$ & $29.3\pm8.2\%$ & $2{,}883\pm269$ & & \\
\bottomrule
\end{tabular}
\end{table}

Table~\ref{tab:phase1} shows Phase~1 results.
DACS significantly outperforms the flat-context baseline at every $N$
($p < 0.0001$, Welch's $t$-test: $t=7.65$ at $N=3$, $t=15.57$ at
$N=5$, $t=34.66$ at $N=10$).
The accuracy gap grows monotonically: $+36.7$ pp ($N=3$), $+58.0$ pp
($N=5$), $+69.0$ pp ($N=10$), while baseline accuracy collapses from
60\% to 21\%.
DACS context grows from 561 to 816 tokens ($+25.5$ tokens/agent,
$R^2>0.99$); baseline grows from 1{,}191 to 2{,}883 tokens.
The efficiency ratio increases at every step: 2.12, 2.72, $3.53\times$.
The elevated DACS contamination at $N=5$ ($14.0\pm10.6\%$, SE) reflects
keyword vocabulary overlap rather than genuine context leakage: Phase~1
answer keywords include generic terms (e.g., \texttt{list}, \texttt{set},
\texttt{deep}) that surface naturally in any technical response regardless
of isolation.
The large SE confirms this is incidental rather than systematic.
Phase~2 scenarios were explicitly designed with disjoint vocabularies,
eliminating this measurement confounder.

\begin{figure}[t]
  \centering
  \includegraphics[width=\textwidth]{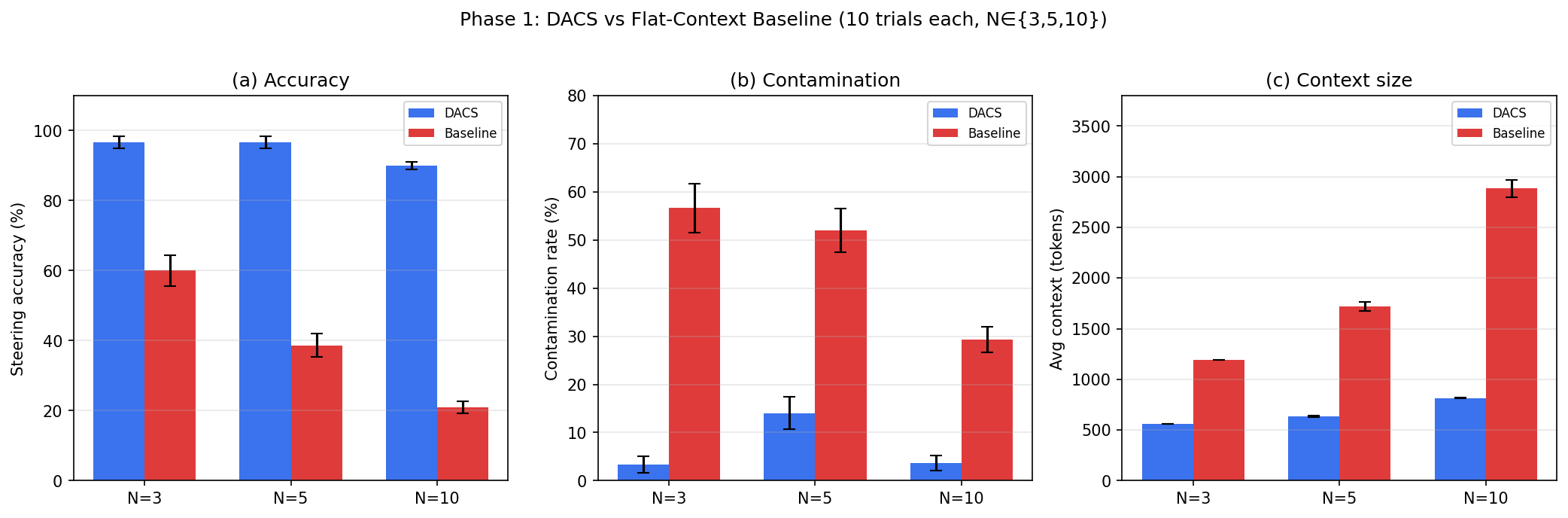}
  \caption{DACS vs.\ flat-context baseline across $N \in \{3,5,10\}$
  (Phase~1).
  Error bars: $\pm 1$ SE (10 trials each).
  (a) Steering accuracy.
  (b) Wrong-agent contamination.
  (c) Average context tokens at steering time.}
  \label{fig:phase1}
\end{figure}

\subsection{Phase 2: Agent Diversity}

\begin{table}[t]
\caption{Phase~2 results across homogeneous, crossfire, and cascade
scenarios.
Each cell: mean $\pm$ SE over 10 trials.}
\label{tab:phase2}
\centering
\footnotesize
\setlength{\tabcolsep}{3pt}
\begin{tabular}{llrrrrr}
\toprule
Scenario & Cond. & Accuracy & Contamination & Ctx (tok) & $\Delta$ acc & Ctx ratio \\
\midrule
\multirow{2}{*}{s4 homogeneous ($N=3$)}
  & DACS     & $90.2\pm3.5\%$ & $0.8\pm0.7\%$  & $815\pm10$  & \multirow{2}{*}{$+37.7$pp} & \multirow{2}{*}{$2.29\times$} \\
  & Baseline & $52.5\pm1.7\%$ & $44.2\pm3.1\%$ & $1{,}869\pm49$ & & \\[4pt]
\multirow{2}{*}{s5 crossfire ($N=5$)}
  & DACS     & $96.0\pm0.9\%$ & $0.0\pm0.0\%$ & $911\pm6$   & \multirow{2}{*}{$+59.0$pp} & \multirow{2}{*}{$2.90\times$} \\
  & Baseline & $37.0\pm1.8\%$ & $53.0\pm2.8\%$ & $2{,}643\pm33$ & & \\[4pt]
\multirow{2}{*}{s6 cascade ($N=5$, adversarial)}
  & DACS     & $94.0\pm1.5\%$ & $7.3\pm2.2\%$  & $705\pm10$  & \multirow{2}{*}{$+37.3$pp} & \multirow{2}{*}{$2.65\times$} \\
  & Baseline & $56.7\pm3.2\%$ & $28.7\pm3.4\%$ & $1{,}870\pm38$ & & \\
\bottomrule
\end{tabular}
\end{table}

Table~\ref{tab:phase2} answers RQ3 (does the advantage generalise
across agent heterogeneity?): DACS wins in all three cases by large
margins ($p < 0.0001$ throughout).

\paragraph{Homogeneous agents (s4).}
Even when all three agents share the same domain vocabulary (algorithm
coding), DACS achieves $+37.7$ pp accuracy.
Same-domain sharing reduces baseline contamination relative to Phase~1
heterogeneous scenarios (44\% vs.\ 57\%), but the flat context's
inability to focus on the current decision point still causes baseline
accuracy to fall to 52.5\%.

\paragraph{Maximum diversity (s5).}
The crossfire scenario is DACS's best case: five domains with
guaranteed vocabulary disjointness.
DACS achieves 96.0\% accuracy with 0.0\% contamination.
The 59.0 pp gap is the second-largest observed across all experiments.
Notably, 5 of 8 disagreements in the LLM-as-judge validation
(\S\ref{sec:metrics}) arise in the baseline condition of this
scenario: contaminated responses that accidentally contain a keyword
from the correct domain while addressing a different agent's task.
This demonstrates that keyword false positives in the baseline
condition are themselves evidence of the contamination mechanism, and
that reported baseline accuracy is if anything a slight overestimate.

\paragraph{Adversarial cascade (s6).}
The cascade scenario is designed to challenge DACS: agents produce
outputs that downstream agents depend on, so the flat baseline might
benefit from knowing all agents' histories.
DACS wins by $+37.3$ pp despite this. The flat context does not
help the baseline orchestrator; it still conflates agent contexts at
steering time.
DACS contamination in s6 (7.3\%) is elevated relative to other
scenarios: the legitimate cross-agent references in a pipeline
(e.g., ``retrieval service decided on BM25'') cause some controlled
vocabulary bleed. Nevertheless, accuracy is dominant.

\begin{figure}[t]
  \centering
  \includegraphics[width=\textwidth]{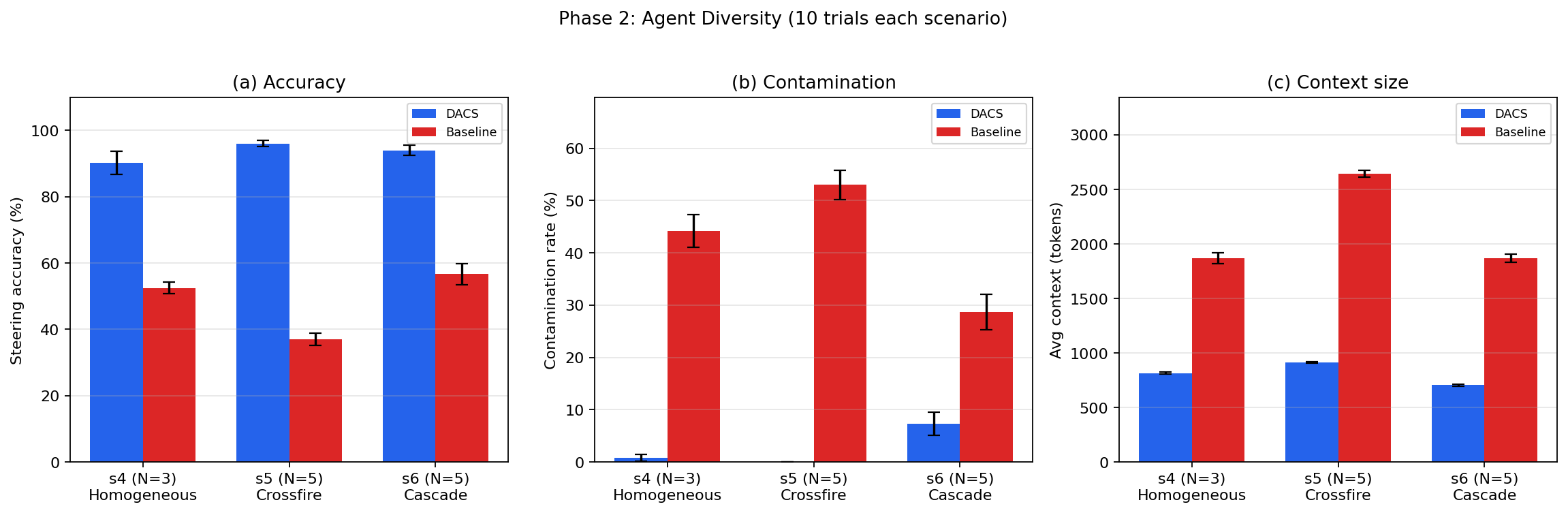}
  \caption{Phase~2 (agent diversity) results across s4 homogeneous,
  s5 crossfire, and s6 cascade.
  Error bars: $\pm 1$ SE (10 trials each).
  (a) Steering accuracy.
  (b) Wrong-agent contamination.
  (c) Average context tokens at steering time.}
  \label{fig:phase2}
\end{figure}

Figure~\ref{fig:phase2} visualises all three metrics for the
heterogeneity scenarios and shows that contamination suppression under
DACS remains robust in both homogeneous and maximally diverse settings.

\subsection{Phase 3: Decision Density Scaling}

\begin{table}[t]
\caption{Phase~3 results: high-density decision scenarios.
$D$ = decisions per agent per trial.
Values are mean $\pm$ SE over 10 trials.}
\label{tab:phase3}
\centering
\footnotesize
\setlength{\tabcolsep}{5pt}
\begin{tabular}{llrrrrr}
\toprule
Scenario & Cond. & Accuracy & Contamination & Ctx (tok) & $\Delta$ acc & Ctx ratio \\
\midrule
\multirow{2}{*}{s7 ($N=5,\; D=8$)}
  & DACS     & $94.0\pm0.8\%$ & $0.2\pm0.2\%$  & $1{,}654\pm20$ & \multirow{2}{*}{$+59.2$pp} & \multirow{2}{*}{$3.24\times$} \\
  & Baseline & $34.8\pm1.8\%$ & $49.8\pm3.3\%$ & $5{,}364\pm98$  & & \\[4pt]
\multirow{2}{*}{s8 ($N=3,\; D=15$)}
  & DACS     & $98.4\pm0.5\%$ & $0.9\pm0.9\%$  & $2{,}755\pm29$ & \multirow{2}{*}{$+54.2$pp} & \multirow{2}{*}{$2.39\times$} \\
  & Baseline & $44.2\pm1.3\%$ & $51.6\pm2.4\%$ & $6{,}573\pm153$ & & \\
\bottomrule
\end{tabular}
\end{table}

Table~\ref{tab:phase3} answers RQ4 (does the advantage scale with
decision density?): yes. Both scenarios show large, highly significant
gains ($t=30.24$, $p=2 \times 10^{-13}$ for s7; $t=40.30$,
$p=10^{-13}$ for s8).

\begin{figure}[t]
  \centering
  \includegraphics[width=\textwidth]{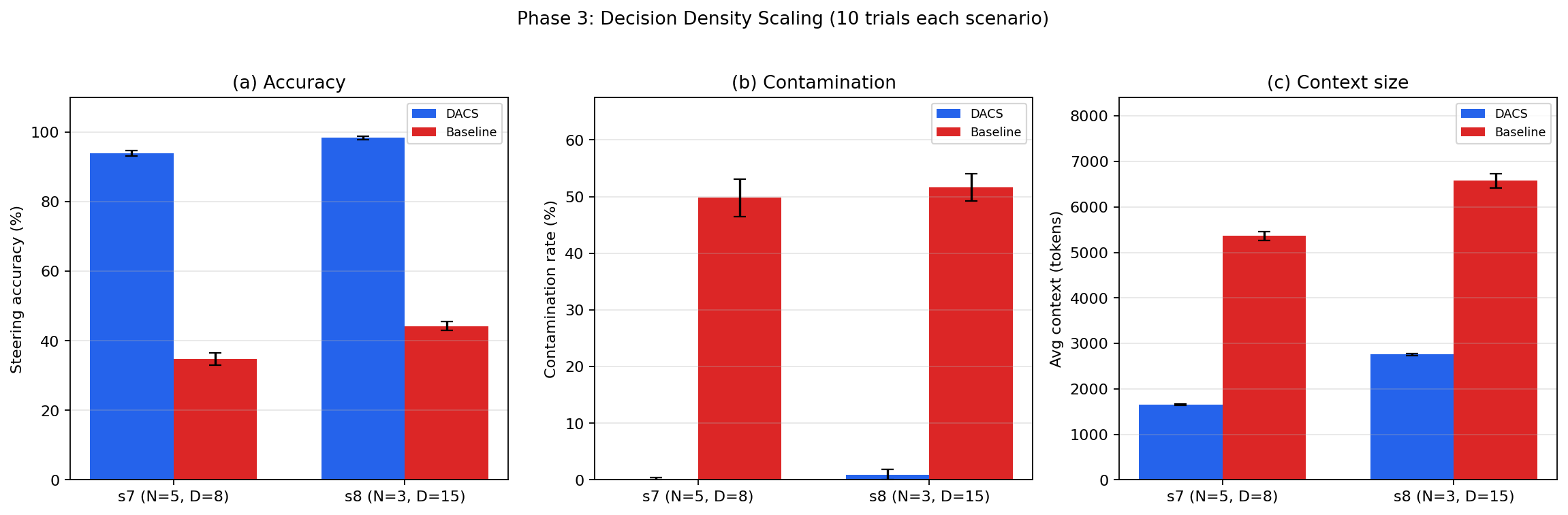}
  \caption{Phase~3 (decision density scaling) results for s7 ($N=5,
  D=8$) and s8 ($N=3, D=15$).
  Error bars: $\pm 1$ SE (10 trials each).
  (a) Steering accuracy.
  (b) Wrong-agent contamination.
  (c) Average context tokens at steering time.}
  \label{fig:phase3}
\end{figure}

Figure~\ref{fig:phase3} shows that high decision density increases
baseline context burden sharply while DACS remains comparatively stable,
preserving both accuracy and contamination control.

\paragraph{RQ4: Decision density amplifies the baseline's degradation.}
Table~\ref{tab:density} tracks the N=3 and N=5 trajectories across all
three phases as $D$ increases.

\begin{table}[t]
\caption{DACS advantage as a function of decisions per agent $D$,
holding $N$ fixed. Accuracy delta and context efficiency ratio grow
as $D$ increases for the baseline; DACS accuracy is stable.}
\label{tab:density}
\centering
\small
\begin{tabular}{lllrrrrr}
\toprule
Phase & Scenario & $N$ & $D$ & DACS acc & Baseline acc & $\Delta$ & Ctx ratio \\
\midrule
1 & s1\_n3           & 3 & $\approx 3$ & 96.7\% & 60.0\% & $+36.7$pp & $2.12\times$ \\
2 & s4\_n3\_homog.   & 3 & 4           & 90.2\% & 52.5\% & $+37.7$pp & $2.29\times$ \\
3 & s8\_n3\_dense\_d3 & 3 & \textbf{15} & \textbf{98.4\%} & \textbf{44.2\%} & $\mathbf{+54.2}$\textbf{pp} & $2.39\times$ \\
\midrule
1 & s2\_n5           & 5 & $\approx 3$ & 96.7\% & 38.7\% & $+58.0$pp & $2.72\times$ \\
2 & s5\_n5\_crossfire & 5 & 4           & 96.0\% & 37.0\% & $+59.0$pp & $2.90\times$ \\
3 & s7\_n5\_dense\_d2 & 5 & \textbf{8}  & \textbf{94.0\%} & \textbf{34.8\%} & $+59.2$pp & $\mathbf{3.24\times}$ \\
\bottomrule
\end{tabular}
\end{table}

At $N=3$, pushing $D$ from 3 to 15 drops baseline accuracy from 60.0\%
to 44.2\%, a 15.8 pp decline, while DACS accuracy rises slightly
(96.7\% $\to$ 98.4\%).
The accuracy delta jumps from $+36.7$ pp to $+54.2$ pp, a 17.5 pp
increase attributable entirely to decision history depth.
At $N=5$, baseline accuracy falls moderately (38.7\% $\to$ 34.8\% as
$D$ doubles), while DACS stays near 94--96\%.
The context efficiency ratio reaches $3.24\times$ for s7, the highest
recorded for $N=5$, confirming that flat-context token costs grow with
accumulated steering history while DACS costs only reflect the current
agent's history.

DACS's accuracy improvement with depth (98.4\% at $D=15$ vs.\ 96.7\%
at $D=3$) is consistent with the mechanism design: at high $D$, the
focus context contains a richer per-agent history, giving the
orchestrator more signal to make accurate steering decisions.
The flat baseline cannot exploit this signal; it is buried under all
other agents' equally long histories.

\subsection{Cumulative View}
\label{sec:cumulative}

\begin{table}[t]
\caption{All 8 scenarios across three phases.
DACS accuracy is consistently high (90.0--98.4\%); baseline collapses
under agent count and decision density pressure.}
\label{tab:all}
\centering
\small
\begin{tabular}{llllrrr}
\toprule
Phase & Scenario & $N$ & $D$ & DACS & Baseline & $\Delta$ \\
\midrule
1 & s1\_n3                & 3  & $\approx$3 & 96.7\% & 60.0\% & $+36.7$pp \\
1 & s2\_n5                & 5  & $\approx$3 & 96.7\% & 38.7\% & $+58.0$pp \\
1 & s3\_n10               & 10 & $\approx$3 & 90.0\% & 21.0\% & $+69.0$pp \\
2 & s4\_n3\_homogeneous   & 3  & 4          & 90.2\% & 52.5\% & $+37.7$pp \\
2 & s5\_n5\_crossfire     & 5  & 4          & 96.0\% & 37.0\% & $+59.0$pp \\
2 & s6\_n5\_cascade       & 5  & 3          & 94.0\% & 56.7\% & $+37.3$pp \\
3 & s7\_n5\_dense\_d2     & 5  & 8          & 94.0\% & 34.8\% & $+59.2$pp \\
3 & s8\_n3\_dense\_d3     & 3  & 15         & 98.4\% & 44.2\% & $+54.2$pp \\
\bottomrule
\end{tabular}
\end{table}

Across all 8 scenarios and 160 trials (Table~\ref{tab:all}), DACS
accuracy ranges $90.0$--$98.4\%$.
The minimum DACS advantage is $+36.7$ pp; the maximum is $+69.0$ pp.
DACS has never lost to the flat-context baseline in any scenario.
The results answer all four research questions:
\textbf{RQ1}~(Does DACS outperform the baseline?): yes, in all 8
scenarios at $p<0.0001$.
\textbf{RQ2}~(Does the advantage grow with $N$?): yes, from
$+36.7$ pp at $N=3$ to $+69.0$ pp at $N=10$.
\textbf{RQ3}~(Does the advantage hold across agent heterogeneity?):
yes, including the adversarial cascade scenario.
\textbf{RQ4}~(Does decision density amplify the advantage?): yes,
$+17.5$ pp increase in the delta as $D$ quintuples at $N=3$.

\section{Phase 4: Real-Agent Validation}
\label{sec:phase4}

\subsection{Motivation}
\label{sec:motivation}

Phases 1--3 use scripted agent stubs: each ``agent'' emits pre-defined
steering questions at pre-defined steps, providing exact experimental
control but raising the obvious question of ecological validity.
Do the results hold when agents are actual LLMs generating their own
reasoning and questions?
Phase~4 answers this with a controlled robustness check: we re-run the
$N{=}3$ and $N{=}5$ scenarios with all stubs replaced by autonomous LLM
agents, keeping the orchestrator, harness, and evaluation protocol
identical.
The goal is not to supersede Phases 1--3 (the scripted harness enables
causal isolation that real agents cannot provide), but to confirm that
the DACS advantage is not an artefact of the controlled evaluation regime.
Two independent experimental paradigms pointing in the same direction are
stronger than one perfect paradigm.

\subsection{Setup}
\label{sec:setup_p4}

\paragraph{Scenarios.}
Two scenarios mirror their Phase~1 counterparts:
\textbf{ra1\_n3} ($N{=}3$: BST implementation, transformer attention survey,
CSV pre-processing) mirrors s1\_n3;
\textbf{ra2\_n5} ($N{=}5$: the above plus graph BFS/DFS/cycle-detection and
RL policy-gradient survey) mirrors s2\_n5.
Both conditions ran 10 DACS and 10 baseline trials (40 total),
with results written to per-trial \texttt{.jsonl} logs as in Phases 1--3.

\paragraph{LLMAgent.}
Each agent runs an independent LLM conversation loop (Claude Haiku~4.5
via OpenRouter; the orchestrator also uses Haiku, keeping a single-model
stack).
Agents emit a \texttt{[[STEER:~\ldots]]} marker whenever they need
orchestrator guidance; the harness intercepts this via regex and routes it
through the standard \textsc{SteeringRequest} protocol, identical to how
scripted stubs inject requests.
Agents may emit at most three steering requests per task, running for at
most 12 LLM steps.
Crucially, question text is the agent's own free-form reasoning:
it is not templated, varies in length and vocabulary across trials, and
may be incomplete or imprecise, a harder evaluation regime than the
scripted keywords admit.
Both scenarios required agents to consult the orchestrator on each
labelled decision category before emitting \texttt{[[DONE]]}, achieving
100\% rubric coverage in both conditions and all 20 trials per scenario.

\paragraph{Evaluation.}
Because questions are free-form, keyword matching alone is insufficient
for the Phase~4 judge.
We use an LLM-as-judge approach with \emph{two independent judge models}:
Claude Haiku~4.5 and GPT-4o-mini.
Each judge receives (i)~the actual question the agent asked,
(ii)~the orchestrator's full response, and
(iii)~a per-rubric \texttt{judge\_context} string describing the correct
decision rationale.
The judge emits a \texttt{<verdict>CORRECT|INCORRECT</verdict>} verdict.
Running two judges allows us to check whether reported effects are
evaluator-specific.
Scenarios ra1\_n3 and ra2\_n5 define 9 and 15 rubric slots per trial
(3 and 5 agents $\times$ 3 rubrics each), yielding 180 and 300 judged
decisions per condition pair across 10 trials.

\subsection{Results}
\label{sec:results_p4}

\begin{table}[t]
\caption{Phase~4 results: real-agent $N$-scaling validation.
Agents are autonomous LLMs (Claude Haiku~4.5) generating free-form
steering questions.
Two independent judges shown.
Phase~1 rows reproduce Table~\ref{tab:phase1} values for cross-paradigm
comparison.
Welch's $t$-test significance on per-trial Haiku-judge accuracy.}
\label{tab:real_agent}
\centering
\small
\begin{tabular}{llrccrrrr}
\toprule
Scenario & $N$ & \multicolumn{2}{c}{M1 Haiku Judge} & \multicolumn{2}{c}{M1 GPT-4o-mini} & \multicolumn{2}{c}{M3 Avg Ctx Tok} & $p$ \\
\cmidrule(lr){3-4}\cmidrule(lr){5-6}\cmidrule(lr){7-8}
 & & DACS & Base & DACS & Base & DACS & Base & \\
\midrule
\multicolumn{9}{l}{\textit{Phase~4 — real agents (LLMAgent, Claude Haiku~4.5, 10 trials each)}} \\
ra1\_n3 & 3 & 79.8\% & 62.6\% & 85.4\% & 67.7\% & 654 & 1{,}361 & 0.0023\,** \\
ra2\_n5 & 5 & \textbf{83.7\%} & 63.3\% & \textbf{89.7\%} & 68.2\% & 799 & 2{,}275 & 0.0008\,*** \\
\midrule
\multicolumn{9}{l}{\textit{Phase~1 synthetic stubs (for comparison; values from Table~\ref{tab:phase1})}} \\
s1\_n3  & 3 & 96.7\% & 60.0\% & --- & --- & 561 & 1{,}191 & $<$0.0001\,*** \\
s2\_n5  & 5 & 96.7\% & 38.7\% & --- & --- & 633 & 1{,}720 & $<$0.0001\,*** \\
\bottomrule
\end{tabular}
\end{table}

\begin{figure}[t]
  \centering
  \includegraphics[width=0.85\textwidth]{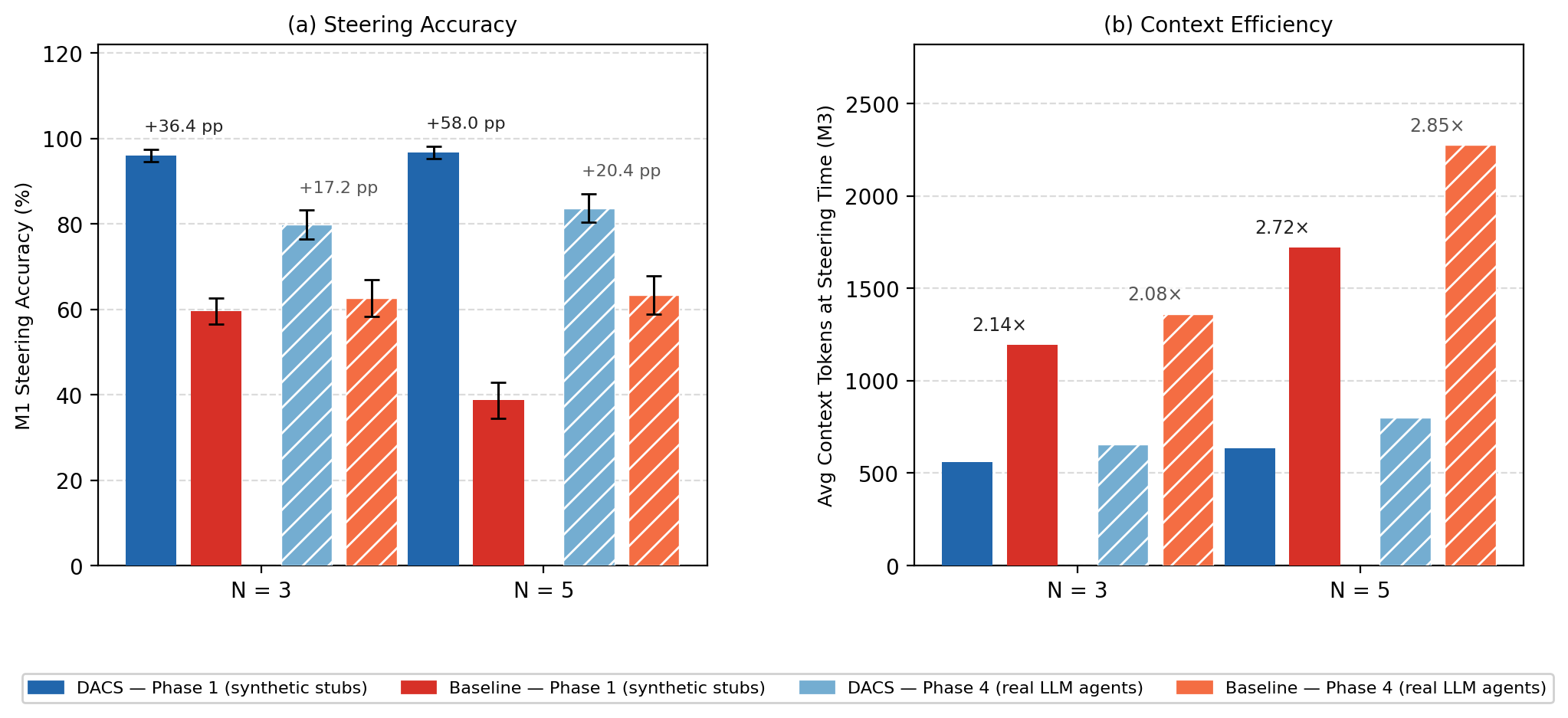}
  \caption{Phase~4 real-agent vs.\ Phase~1 synthetic results at matched
  $N \in \{3, 5\}$.
  (a)~M1 steering accuracy (Haiku judge) for DACS and baseline across
  both paradigms.
  (b)~M3 average context tokens at steering time.
  Error bars: $\pm 1$ SE (10 trials per condition).
  DACS advantage direction and growth with $N$ are consistent across
  paradigms; absolute accuracy is lower under real agents due to the
  harder free-form question regime.}
  \label{fig:phase4}
\end{figure}

Table~\ref{tab:real_agent} and Figure~\ref{fig:phase4} show four
findings that replicate and extend the Phases 1--3 results.

\noindent\textbf{(F1) DACS significantly outperforms the baseline in
both real-agent scenarios.}
The Haiku judge records $+17.2$\,pp at $N{=}3$ ($p{=}0.0023$) and
$+20.4$\,pp at $N{=}5$ ($p{=}0.0008$).
Both results are independently confirmed by GPT-4o-mini ($+17.7$\,pp and
$+21.5$\,pp respectively), ruling out evaluator-specific artefacts.

\noindent\textbf{(F2) The advantage grows with $N$, consistent with
Phase~1.}
At $N{=}3$ DACS leads by $+17.2$\,pp; at $N{=}5$ by $+20.4$\,pp.
The same monotonic growth was observed in Phase~1
($+36.7$\,pp at $N{=}3$, $+58.0$\,pp at $N{=}5$, $+69.0$\,pp at
$N{=}10$), confirming that context isolation becomes more valuable
as the number of concurrent agents grows, regardless of whether those
agents are scripted or autonomous.

\noindent\textbf{(F3) Context efficiency is preserved.}
DACS context at $N{=}3$ is $654$ tokens vs.\ $1{,}361$ baseline
($2.08\times$); at $N{=}5$, $799$ vs.\ $2{,}275$ ($2.85\times$).
The ratio grows with $N$, again consistent with Phases 1--3.

\noindent\textbf{(F4) Absolute accuracy under real agents is lower,
as expected.}
Real-agent DACS reaches 79.8--83.7\% vs.\ 96.0--96.7\% for scripted
stubs.
This is the expected cost of the harder regime: agents produce irregular,
sometimes incomplete questions that do not always map cleanly to rubrics,
and keyword matching at 99.2\% (DACS) is not the binding constraint.
The LLM judges capture these partial answers, lowering recorded accuracy
relative to the synthetic benchmark.
Critically, the \emph{relative} advantage is present and significant in
both paradigms.

\subsection{The Contamination Gap}
\label{sec:contam_gap}

The most striking divergence between Phase~4 and Phases~1--3 is in M2
contamination.
In the synthetic experiments DACS contamination is near-zero
($0.0$--$14.0\%$, the upper end driven entirely by vocabulary overlap
artefacts).
In Phase~4, DACS contamination is $59.9\%$ at $N{=}3$ and $40.3\%$ at
$N{=}5$.

The explanation is the nature of real agent outputs.
During a \textsc{Focus}$(a_i)$ session the registry entries for all other
agents are present as compact summaries (e.g.,
``\texttt{a2: RUNNING, transformer survey, 3/5 steps done}'').
A verbose model like Haiku tends to contextualise its advice using these
registry labels: ``\emph{unlike a2's retrieval task, your BST inorder walk
benefits from\ldots}''
This cross-reference is a \emph{helpful} grounding behaviour in a real
system: the orchestrator is showing it understands which agent it is
addressing, but it trips the M2 binary detector, which flags any response
containing another agent's ID.
This is a measurement artefact, not a fidelity failure.

Evidence: under DACS, M1 (decision quality) is significantly higher than
the baseline despite the elevated M2 reading.
Under the baseline, M2 approaches ceiling in both scenarios (95.8\%),
compared to 54.5\% and 52.0\% for matching synthetic scenarios, a
$+41$--$44$\,pp increase consistent with the baseline's flat-context
exposure causing genuine cross-agent vocabulary bleed into every response,
not just registry citations.

This gap points to two directions for future work.
First, the DACS \textsc{SteeringRequest} format could be tightened:
requiring agents to submit structured requests (``agent\_id'',
``decision\_type'', ``question'') rather than free text would constrain
how the orchestrator contextualises its response, reducing
incidental cross-references.
Second, a richer contamination metric (e.g., tracking whether
cross-agent vocabulary appears in the \emph{decision clause} of the
response rather than in framing or context) would separate harmful
leakage from helpful grounding in real-agent deployments.

\section{Discussion}
\label{sec:discussion}

\subsection{Why DACS Works}

The mechanism's effectiveness follows directly from what it removes from
the context window.
In the flat baseline, when $a_i$ asks ``should I use recursive or
iterative traversal for my BST inorder walk?'', the context simultaneously
holds a research agent's detailed transformer attention analysis and a
data processor's CSV encoding problem, containing domain keywords
(``attention'', ``heads'', ``encoding'', ``UTF-8'') that have nothing to
do with the correct answer.
The LLM anchors its response to whichever content pattern captures its
attention, an effect we measure as both inaccuracy and contamination.

DACS removes this anchor competition entirely.
In \textsc{Focus}$(a_i)$ mode, the context contains $a_i$'s task
description and steering history plus compact status lines for other
agents (``a2: RUNNING, transformer survey, 3/5 steps done'').
The correct-answer vocabulary dominates the context by construction.

\subsection{Why High Decision Density Hurts the Baseline}

Phase~3 reveals a compounding mechanism.
As $D$ grows, the flat context accumulates not just $N$ agents'
\emph{current} states but their full steering \emph{histories}.
At $D=15$, a 3-agent flat context holds 45 steering interaction pairs,
each injecting domain vocabulary that drowns out the signal for the
current decision.
DACS's focus context at $D=15$ holds only $a_i$'s 15-interaction
history (precisely the relevant signal) plus three compact registry
lines for the other agents.
The result is that DACS accuracy at $D=15$ (98.4\%) \emph{exceeds} its
accuracy at $D\approx 3$ (96.7\%): more history in the focus
context is additive signal, not noise.

\subsection{Relation to Hierarchical Routing}

DACS is orthogonal to hierarchical agent architectures
(AgentOrchestra~\citep{agentorchestra2025},
AdaptOrch~\citep{adaptorch2025b}).
Hierarchical routing reduces how many agent threads reach the
orchestrator at all.
DACS controls what the orchestrator holds in its context at each
steering moment during execution.
A production system could combine both: hierarchical routing limits the
active agent pool, and DACS ensures that within that pool, each steering
interaction is isolated.

\subsection{Limitations}

\paragraph{Benchmark scope.}
The task suite uses keyword matching to evaluate steering accuracy.
This is conservative (multi-word phrases and paraphrases are not captured)
and may produce false-positive contamination detections when domains share
vocabulary.
LLM-as-judge validation across Phases~2 and~3 confirms metric validity:
Phase~2 $\kappa=0.956$; Phase~3 s8 $\kappa=0.886$; Phase~3 s7
$\kappa=0.933$ (mean $\kappa=0.909$, all well above the $\kappa{\geq}0.80$
``substantial agreement'' threshold).
Phase~4 uses LLM judges as the primary metric throughout, avoiding
keyword matching for free-form agent questions entirely.

\paragraph{Scripted agent scope (mitigated by Phase~4).}
Phases~1--3 agent stubs simulate decision points with pre-defined timing
and question templates.
Phase~4 (\S\ref{sec:phase4}) directly addresses this: all stubs are
replaced by autonomous LLM agents (Claude Haiku~4.5), and the M1 accuracy
advantage replicates in both conditions (DACS $+17.2$\,pp at $N{=}3$,
$p{=}0.0023$; $+20.4$\,pp at $N{=}5$, $p{=}0.0008$; confirmed by
GPT-4o-mini), confirming that DACS's benefits are not an artefact of
the scripted harness.
The remaining gap is that Phase~4 covers only $N{\in}\{3,5\}$ at low
decision density ($D{\approx}3$) and one agent model family (Haiku).
Generalisation to $N=10$, high $D$, and other model families remains
open future work.
M2 contamination is less discriminative with real agents because
chatty LLMs reference registry-provided agent IDs in focus-mode
responses (see \S\ref{sec:contam_gap}); M1 remains the primary quality
indicator for real-agent deployments.

\paragraph{Model coverage.}
Phases~1--3 use MiniMax-M2.7 (context window 204{,}800 tokens);
Phase~4 uses Claude Haiku~4.5 via OpenRouter.
The consistent DACS advantage across two distinct model families provides
initial evidence of model-family independence, but generalisation to
smaller context budgets, instruction-tuned models with different attention
characteristics, or proprietary frontier models (GPT-4o, Claude Opus)
has not been tested.

\paragraph{Interrupt handling not experimentally isolated.}
The interrupt protocol (transition~\eqref{eq:t3}) was exercised naturally in
experiments but not ablated independently.

\subsection{Future Work}

\begin{itemize}
  \item \textbf{Full LLM-as-judge pass for Phase~3.} The current judge
  validation used stratified samples (100 decisions from s8, 200 from s7).
  A full pass over all 900 decisions in s8\_n3\_dense\_d3 would provide
  complete coverage and tighter $\kappa$ estimates.
  \item \textbf{Phase~4 at $N=10$ and high $D$.} Phase~4 covers only
  $N{\in}\{3,5\}$ at $D{\approx}3$, mirroring Phase~1.
  Extending to $N=10$ and to the high-density regime ($D{\in}\{8,15\}$)
  would close the remaining ecological-validity gap for the full
  Phase~1--3 design space.
  \item \textbf{Contamination metric refinement.} The binary agent-ID
  mention detector (M2) conflates helpful registry grounding
  (``unlike a2's task\ldots'') with genuine decision-level leakage.
  A clause-level metric that fires only when cross-agent vocabulary
  appears in the decision clause, not the framing, would provide a
  cleaner real-agent contamination signal.
  Tighter \textsc{SteeringRequest} formatting (structured fields rather
  than free text) would also reduce incidental cross-references.
  \item \textbf{User responsiveness.} Measure latency from user message
  to orchestrator response under DACS vs.\ baseline at varying $N$ and
  queue depth.
  \item \textbf{Very high $D$.} Extend Phase~3 to $D \in \{30, 50, 100\}$
  to probe where, if anywhere, DACS accuracy degrades as per-agent
  history grows very large.
  \item \textbf{Cross-model generalisation.} Test with agent and
  orchestrator models beyond Haiku and MiniMax-M2.7 (e.g., GPT-4o-mini
  as agent, Claude Opus as orchestrator) to probe whether the advantage
  holds under different attention and instruction-following profiles.
\end{itemize}

\section{Conclusion}
\label{sec:conclusion}

We presented DACS, a dynamic attentional context scoping mechanism that
solves context pollution in multi-agent LLM orchestration through
agent-triggered asymmetric focus sessions.
The mechanism is exact (not approximate), sub-linear in agent count, and
compatible with existing orchestration architectures.

Across 160 synthetic trials in 8 scenarios spanning three phases,
DACS achieves $90.0$--$98.4\%$ steering accuracy versus $21.0$--$60.0\%$
for the flat-context baseline, reduces wrong-agent contamination from
$28$--$57\%$ to $0$--$14\%$, and exhibits growing context efficiency
(2.12--3.53$\times$) as $N$ increases.
The advantage is robust to agent heterogeneity (Phase~2), holds even in
pipeline-dependent adversarial scenarios, and \emph{compounds with
decision density} (Phase~3): quintupling $D$ from 3 to 15 causes baseline
accuracy to fall by 15.8~pp while DACS accuracy is unchanged or improves.
Phase~4 validates the core result beyond the scripted benchmark:
two real-agent experiments (ra1\_n3 and ra2\_n5, 40 trials total,
10 DACS $+$ 10 baseline each) replace all stubs with autonomous LLM
agents (Claude Haiku~4.5) generating free-form steering questions:
DACS 79.8\% vs.\ baseline 62.6\% at $N{=}3$ ($+17.2$~pp, $p{=}0.0023$);
DACS 83.7\% vs.\ baseline 63.3\% at $N{=}5$ ($+20.4$~pp, $p{=}0.0008$).
Both results are independently confirmed by GPT-4o-mini.
The advantage grows with $N$ in the real-agent setting ($+17.2 \to +20.4$~pp),
mirroring the synthetic scaling trend.
Absolute accuracy is lower than with scripted stubs, reflecting the
harder free-form regime, but the direction and significance of the
DACS effect are unambiguous across both experimental paradigms.

The core contribution is demonstrating that \emph{what is in the
orchestrator's context window at the moment of steering} is the
dominant variable for multi-agent steering accuracy, not model size,
not topology, not memory compression, and that a simple, deterministic
mode-switch mechanism suffices to control it effectively across a
wide range of agent counts, diversity structures, and decision densities.

\bibliographystyle{plainnat}
\bibliography{refs}

@article{afm2024,
  title   = {Adaptive Focus Memory for Language Models},
  author  = {Cruz, Christopher},
  journal = {arXiv preprint arXiv:2511.12712},
  year    = {2024},
  url     = {https://arxiv.org/abs/2511.12712}
}

@article{aoi2024,
  title   = {{AOI}: Context-Aware Multi-Agent Operations via Dynamic Scheduling and Hierarchical Memory Compression},
  author  = {Bai, Zishan and Luo, Jing and Ni, Ziyi and Ge, Enze and Shi, Jiacheng and Zhang, Yichao and Gu, Jiayi and Han, Zhimo and Bao, Riyang and others},
  journal = {arXiv preprint arXiv:2512.13956},
  year    = {2024},
  url     = {https://arxiv.org/abs/2512.13956}
}

@article{agentorchestra2025,
  title   = {{AgentOrchestra}: Orchestrating Multi-Agent Intelligence with the Tool-Environment-Agent ({TEA}) Protocol},
  author  = {Zhang, Wentao and Zeng, Liang and Xiao, Yuzhen and Li, Yongcong and Cui, Ce and Zhao, Yilei and Hu, Rui and Liu, Yang and Zhou, Yahui and others},
  journal = {arXiv preprint arXiv:2506.12508},
  year    = {2025},
  url     = {https://arxiv.org/abs/2506.12508}
}

@article{ace2026,
  title   = {{ACE}: Agentic Context Engineering: Evolving Contexts for Self-Improving Language Models},
  author  = {Zhang, Qizheng and Hu, Changran and Upasani, Shubhangi and Ma, Boyuan and Hong, Fenglu and Kamanuru, Vamsidhar and Rainton, Jay and Wu, Chen and others},
  journal = {arXiv preprint arXiv:2510.04618},
  year    = {2026},
  url     = {https://arxiv.org/abs/2510.04618}
}

@article{adaptorch2025b,
  title   = {{AdaptOrch}: Task-Adaptive Multi-Agent Orchestration in the Era of {LLM} Performance Convergence},
  author  = {Yu, Geunbin},
  journal = {arXiv preprint arXiv:2602.16873},
  year    = {2025},
  url     = {https://arxiv.org/abs/2602.16873}
}

@article{codedelegator2025,
  title   = {{CodeDelegator}: Mitigating Context Pollution via Role Separation in Code-as-Action Agents},
  author  = {Fei, Tianxiang and Chen, Cheng and Pan, Yue and Zheng, Mao and Song, Mingyang},
  journal = {arXiv preprint arXiv:2601.14914},
  year    = {2025},
  url     = {https://arxiv.org/abs/2601.14914}
}

@article{sidequest2025,
  title   = {{SideQuest}: Model-Driven {KV} Cache Management for Long-Horizon Agentic Reasoning},
  author  = {Kariyappa, Sanjay and Suh, G. Edward},
  journal = {arXiv preprint arXiv:2602.22603},
  year    = {2025},
  url     = {https://arxiv.org/abs/2602.22603}
}

@article{adaptorch2025,
  title   = {Adaptive Orchestration: Scalable Self-Evolving Multi-Agent Systems},
  author  = {Sampath, Sathish and Baskaran, Anuradha},
  journal = {arXiv preprint arXiv:2601.09742},
  year    = {2025},
  url     = {https://arxiv.org/abs/2601.09742}
}

@article{lemon2025,
  title   = {Lemon Agent Technical Report},
  author  = {Jiang, Haipeng and Ren, Kailong and Yin, Zimo and Sun, Zhetao and Gan, Xin and Lv, Guangyi and He, Ming and Wang, Peng and Yin, Congli and Pan, Hong and others},
  journal = {arXiv preprint arXiv:2602.07092},
  year    = {2025},
  url     = {https://arxiv.org/abs/2602.07092}
}

@misc{anthropic2025code,
  title  = {Claude Code Agent Teams},
  author = {{Anthropic}},
  year   = {2025},
  url    = {https://www.anthropic.com/claude-code}
}

@misc{opencode2025,
  title  = {{OpenCode}: The Open Source Coding Agent},
  author = {{Anomaly}},
  year   = {2025},
  url    = {https://github.com/anomalyco/opencode}
}

\end{document}